\documentclass[%
 reprint,
superscriptaddress,
 amsmath,amssymb,
 aps,
]{revtex4-2}
\usepackage{xcolor}
\usepackage{upgreek}
\usepackage{graphicx}
\usepackage{dcolumn}
\usepackage{bm}
\DeclareSymbolFont{matha}{OML}{txmi}{m}{it}
\DeclareMathSymbol{\varv}{\mathord}{matha}{118}
\usepackage{natbib}

\usepackage{lipsum}
\emergencystretch=\maxdimen
\begin{document}

\preprint{APS/123-QED}

\title{Collapse of the vacuum in hexagonal graphene quantum dots: a comparative study between
the tight-binding and the mean-field Hubbard models}
\author{Mustafa Polat}
\affiliation{Izmir Institute of Technology, Department of Physics, 35430 Urla, Izmir, Turkey}%
\email{mustafapolat@iyte.edu.tr}
\author{H$\hat{\text{a}}$ldun Sevin\c{c}li}
\affiliation{Izmir Institute of Technology, Department of Materials Science and Engineering, 35430 Urla, Izmir, Turkey}%
\author{A. D. G\"{u}\c{c}l\"{u}}
\affiliation{Izmir Institute of Technology, Department of Physics, 35430 Urla, Izmir, Turkey}%
\date{\today}
\begin{abstract}
In this paper, we perform a systematic study on the electronic, magnetic, and transport properties
of the hexagonal graphene quantum dots (GQDs) with armchair edges in the presence of a charged impurity using
two different configurations: (1) a central Coulomb potential and (2) a positively charged carbon vacancy.
The tight binding (TB) and the half-filled extended Hubbard models are numerically solved and
compared with each other in order to reveal the effect of electron interactions and system sizes. Numerical results
point out that off-site Coulomb repulsion leads to an increase in the critical coupling constant to $\beta_{\text{c}}$
= 0.6 for a central Coulomb potential. This critical value of the $\beta$ is found to be independent of the GQD size,
reflecting its universality even in the presence of electron-electron interactions. In addition, a sudden downshift in the transmission peaks shows a clear signature of the transition from subcritical $\beta$ $<$ $\beta_{\text{c}}$ to supercritical
$\beta$ $>$ $\beta_{\text{c}}$ regime. On the other hand, for a positively charged vacancy,
collapse of the lowest bound state occurs at $\beta_{\text{c}}$ = 0.7 for the interacting case.
Interestingly, the local magnetic moment, induced by a bare carbon
vacancy, is totally quenched when the vacancy is subcritically charged, whereas the valley splittings in
electron and hole channels continue to exist in both regimes.
\end{abstract}

\maketitle

\section{\label{sec:level1}INTRODUCTION}
The exact solution of 3D Dirac equation in an external Coulomb field,
produced by a point nucleus, is only consistent up to a critical threshold Z$_{\text{c}}$ = $\alpha^{-1}$ $\sim$ 137, where
$\alpha$ = $\emph{e}^2$/$\hbar$\emph{c} is the Sommerfeld fine-structure constant \cite{zeldovich1972electronic}.
For larger values of the nuclear charge Z, the energy eigenvalues become purely imaginary, the wave function
is non-normalizable, and its real part exhibits oscillatory
behavior \cite{greiner1990relativistic}. Actually, the
singularity of the point nucleus at the center leads to a non-self-adjoint Hamiltonian that
could not be properly solved unless a finite-size
for the nucleus is introduced \cite{pomeranchuk1945energy}. This regularization results in a larger critical
threshold of Z$_{c}$ $\sim$ 172 above which the wave function
becomes a narrow resonance with a finite lifetime in compliance
with Fano$^{'}$s formalism \cite{reinhardt1977quantum}. In particular, the lowest bound state 1S$_{1/2}$ with the total
angular momentum quantum number \emph{j} = 1/2 dives into
the negative continuum for a coupling constant $\beta$ = Z$\alpha$
if it exceeds 1, but the direct evidence of such a particular behavior has so far
remained elusive in high energy heavy-ion collision experiments \cite{schweppe1983observation,cowan1985anomalous}.
However, the situation is slightly different in 2+1 dimensions for which the
critical coupling constant $\beta_{\text{c}}$ becomes 1/2 \cite{khalilov1998dirac}.
In this sense, bulk graphene with a larger fine structure constant
$\alpha_{g}$ = 2.2/$\kappa$, where
$\kappa$ is the dielectric constant, could provide an ideal platform in
theory \cite{neto2009electronic}.
Unlike QED, Z$_{c}$ is expected to be on the order of the unity
\cite{shytov2007atomic,terekhov2008screening}, which carries the signs
of experimental transition to the supercritical regime
in a table-top experiment. Indeed, the formation of an infinite family of quasi-bound states
in the presence of the clusters of charged calcium (Ca) dimers on graphene have been successfully monitored
via the local density of states (LDOS) in an experimental study \cite{wang2013observing}.

Later, Mao et al. \cite{mao2016realization} demonstrated that a positive charge can be deposited
into a single carbon vacancy by applying voltage pulses of 2 $-$ 3 V for $>$ 10
seconds with the help of a scanning tunneling microscope (STM) tip. A charged vacancy in graphene is in analogy with
the piling up positively charged ions and similarly leads to the sudden appearance
of a sequence of quasi$-$bound states \cite{mao2016realization}.
Besides, it is well-known that the removal of a single carbon atom lifts the spin
and valley degrees of freedom \cite{neto2009electronic}, hence the local
magnetic moment is induced \cite{yazyev}. Since only a prominent resonant peak
is observed in previous studies \cite{ugeda2010missing,ugeda2011point},
the spin splitting has recently attracted attention in experiments \cite{PhysRevLett,gonzalez2016atomic}.
In addition, for a while there has been significant progress in measurement of the valley
splittings around a carbon vacancy \cite{li} thanks to discrete energy levels and an unconventional method of preparation of
graphene quantum dots (GQDs) \cite{freg,freitag2018large}. The question arises as to what sort
of changes in physical properties happen after a bare vacancy
is positively charged with the subcritical and supercritical Coulomb potentials.

Of all the GQDs that have been reported so far \cite{ezawa2007metallic,fernandez2007magnetism,wang2008graphene,akola2008edge,
schnez2008analytic,mueller2010triplet,wimmer2010robustness,zarenia,hamalainen2011quantum,olle2012yield,
sheng,subramaniam2012wave,szalowski2013indirect,devrim}, the hexagonal GQDs with armchair edges deserve
attention due to the specific band gap feature. Free of
localized edge states, the band gap is proportional to
the inverse square root of number of atoms (E$_{\text{gap}}$ $\propto$ \emph{k}$_{\text{min}}$ $\approx$ 2$\pi$/$\Delta$\emph{x} $\propto$ 1/$\sqrt{\text{N}}$) \cite{sheng}. It corresponds to linear
photon dispersion relation for confined Dirac fermions \cite{gucclu2010excitonic}.
In addition, the hexagonal shaped GQDs have well-known
properties, among which, (1) sublattice symmetry
results in spin symmetry; (2) two doubly degenerate
levels in the vicinity of Fermi level account for the valley
symmetry \cite{devrim}. These internal properties indicate that the pristine hexagonal
GQDs with armchair edges carry all the symmetries of
graphene. Therefore, it becomes possible to follow the evolution of the
spin and valley splittings as a function of the coupling
constant $\beta$ when a point vacancy is deliberately
created and gradually charged. In this work, we construct a set of Dirac vacuums
with the help of the hexagonal GQDs that differ in
size. The tight-binding (TB) and the extended mean-field
Hubbard (MFH) models are separately solved for
the central Coulomb potential and a charged vacancy.
The central impurity in a GQD was previously investigated
using effective mass approach with appropriate boundary conditions \cite{van2017graphene}, but
Hubbard descriptions including electron-electron interactions and spin effects are still missing.

The rest of this paper is organized as follows. In Sec.~\ref{sec:level2} Hamiltonian of the extended MFH model,
the Coulomb potential, and the non-equilibrium Green function formalism are introduced.
The finite-size effect, the effect of electron-electron interactions, and the transmission coefficient
of the lowest bound states for the central potential
are discussed in greater detail in Sec.~\ref{sec:level3}. Both electronic and magnetic properties,
as well as the transmission coefficients are studied
in the presence of a charged vacancy in  Sec.~\ref{sec:level6}.
Finally, Sec.~\ref{sec:level10} consists of our conclusions.
\section{\label{sec:level2} COMPUTATIONAL METHODS}
We solve the extended mean-field theory of the
Hubbard model starting from a single-band tight-binding
approximation for the $\pi_{z}$ orbitals. The single-valley Dirac
description of the $\pi_{z}$ electron dynamics are described by
the following effective Hamiltonian
\begin{align}\label{eq:eq1}
\text{H$_{MFH}$} \nonumber & =t\sideset{}{}\sum_{<ij>\sigma}(c_{i\sigma}^{\dagger}c_{j\sigma}+\text{H.c.})\\
        \nonumber&+\text{U}\sideset{}{}\sum_{i\sigma}(<n_{i\overline{\sigma}}>-\frac{1}{2})n_{i\sigma}\\
        \nonumber & + \sum_{ij}\text{V$_{ij}$}(<n_{j}>-1)n_{i}\\
        & -\hbar \varv_{\text{F}} \beta\sum_{i\sigma}\frac{c_{i\sigma}^{\dagger}c_{i\sigma}}{r_{i}}\text{.}
\end{align}
The operator c$_{i\sigma}^{\dagger}$ (c$_{j\sigma}$) given in the first term describes
the creation (annihilation) of an electron with spin $\sigma$
at the lattice sites. The nearest neighbor hopping term t = -2.8 eV is used, and which preserves the electron-hole
symmetry in the absence of the Coulomb potential \cite{pereira2007coulomb}.
The second term is the on-site interaction term in which
only two electrons with opposite spin can occupy the same lattice site by paying an extra correlation energy of U.
It is taken to be 16.52/$\kappa$ eV, where the dielectric constant $\kappa$ equals to 6
corresponding to the interband polarization \cite{devrim,ando2006screening}. In Eq. \ref{eq:eq1},
the terms $<$n$_{i\sigma}$$>$ and n$_{i\sigma}$ are associated with the spin dependent expectation value
of electron densities and the number operator for an electron
with spin $\sigma$ at the site i, respectively. The off-site Coulomb repulsion on top of the on-site repulsion
is added to Hamiltonian by means of V$_{ij}$ term which is set to be as
8.64/$\kappa$ eV, 5.33/$\kappa$ eV, and 1/$\kappa$d$_{ij}$ eV for the first neighbors, second neighbors, and the
atomic sites at relatively large distances, respectively \cite{devrim,potasz2010spin}. The last term stands for the Coulomb
potential in which r$_{i}$ is the distance between the lattice site i
and the center of the Coulomb potential \cite{moldovan2016atomic}, and $\varv_{\text{F}}$ is the Fermi velocity.
The coupling constant $\beta$ is assumed to be attractive ($>$ 0) without loss of
generality in this study.

As a measurable feature of the critical states, LDOS is obtained by the formula
\begin{align}\label{eq:eq2}
\text{N(E,r)} = \sum_{n}|\Psi_{n}(r)|^{2}\delta(\text{E}-\text{E}_{n})\text{.}
\end{align}
LDOS is projected onto the lattice sites to demonstrate spatial
distribution of the collapse states at different coupling constants.
In the presence of a single charged vacancy, starting from
the self-consistent expectation values
of electron densities, we compute the spin density per lattice site as follows
\begin{align}\label{eq:eq3}
<s_{i}^{z}> = m_{i} = (<n_{i\uparrow}>-<n_{i\downarrow}>)/2 \text{,}
\end{align}
where $<$n$_{i\sigma}$$>$'s are calculated by summing up all states lying below Fermi level.
Starting from Eq. \ref{eq:eq3}, the staggered magnetization as an order parameter of
the antiferromagnetism is numerically calculated from
\begin{align}\label{eq:eq4}
\mu_{s}^{z} = \sum_{i}(-1)^{i}<s_{i}^{z}>\text{,}
\end{align}
where (-1)$^{i}$ indicates that the contributions are summed up from the opposite sublattices
with opposite signs. $\mu_{s}^{z}$ is proportional to the antiferromagnetic
phase \cite{grujic2013antiferromagnetism}.

To calculate the transmission coefficients, we utilize the non-equilibrium Green function (NEGF) formalism.
The transmission coefficients are obtained from

\begin{equation}\label{eq:eq5}
\text{T(E)}=\text{Tr}(\Gamma^{L}(\text{E})\text{G(E)}\Gamma^{R}(\text{E})\text{G}^{\text{T}}(\text{E}))\text{,}
\end{equation}
where

\begin{equation}\label{eq:eq6}
\text{G(E)}=((\text{E}+i0^{+})\text{I}_{\text{N}\times \text{N}}-\text{H}_{c}-\Sigma_{L}-\Sigma_{R})^{-1}
\end{equation}
is the Green function in which $0^{+}$ is 10$^{-6}\times$\emph{t},
H$_{c}$ represents the central Hamiltonian of the analyzed structures, and $\Sigma$$_{L}$ ($\Sigma$$_{R}$) is the self energy matrix
of the left (right) probe, where generic electrodes are used in order to avoid structural features arising from the
electrodes in the resulting transmission spectra. For that purpose, a one-dimensional wide
bandwidth tight-binding chain is assumed. Self energies matrices ($\Sigma$$_{N \times N}$) for the
right and left leads are obtained from the analytical solution
of surface Green function \cite{muller2000understanding}. The probes are placed at the ends of the GQDs and the hopping
term is taken as t/4. In Eq. \ref{eq:eq5}, $\Gamma$$^{L,R}$'s are the corresponding broadening matrices,
and the hopping parameter of \emph{t} in the reservoirs is used \cite{muller2000understanding}.
The transmission coefficients around the resonance energies of the defect-induced and atomic collapse
states are numerically calculated for different values of the $\beta$.
\section{\label{sec:level3} CENTRAL COULOMB POTENTIAL}
\subsection{\label{sec:level4} Size quantization and electron-electron interactions}
\begin{figure}
\begin{center}
\includegraphics[width=\linewidth, height=\textheight,keepaspectratio]{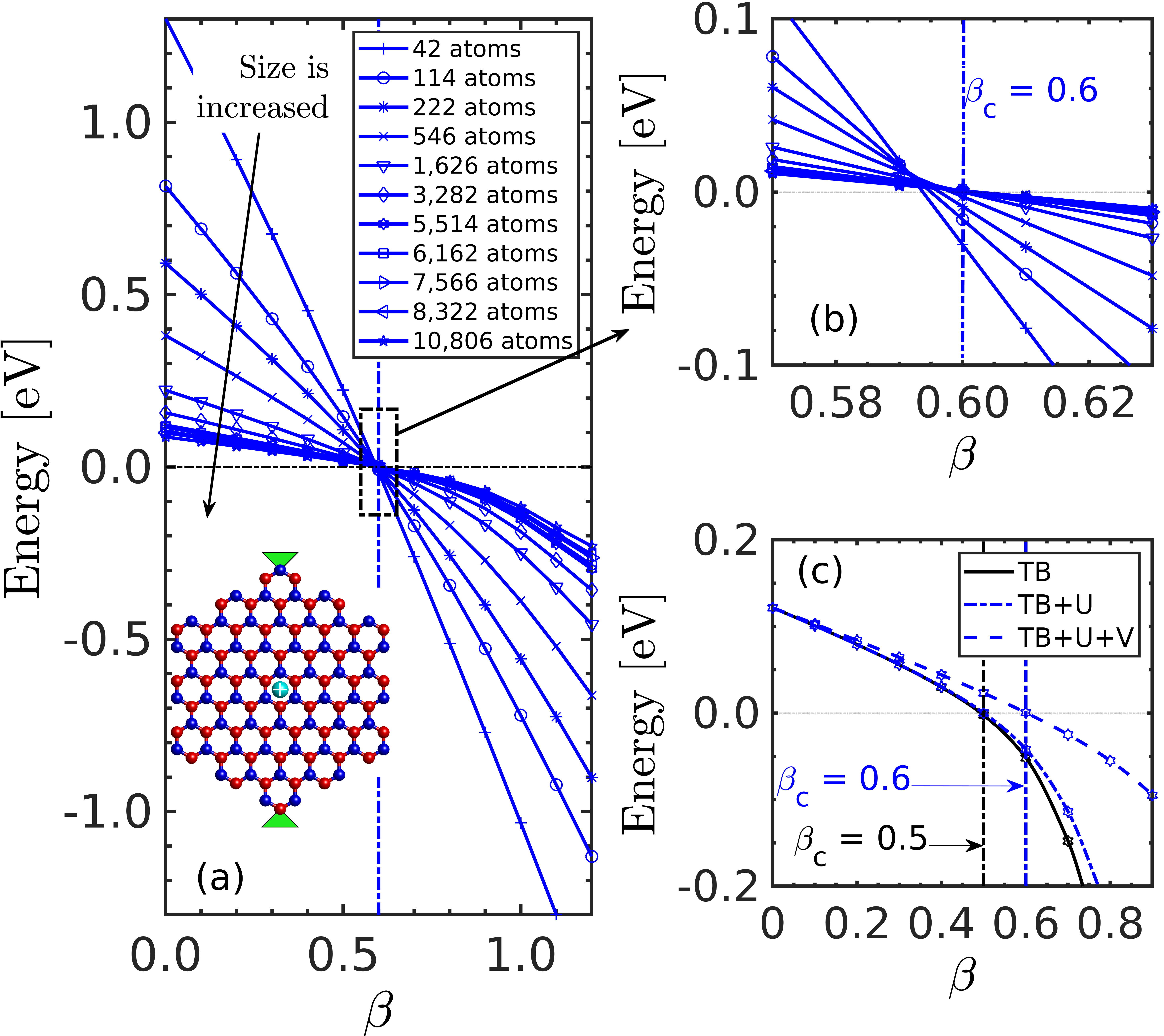}
\caption{\label{fig:fig_1} Energy values of the lowest bound states as a function of the coupling constant $\beta$.
(a) The critical coupling constant $\beta_{c}$ is 0.6 within the the MFH model for all samples that differ in size.
The inset contains a sketch of the problem for the hexagonal GQD that consists of 114 atoms. Here the sublattices A and B are red and blue filled circles, and a positively charged impurity is at the center. Green triangles show how the leads are connected to samples throughout our study to determine the transmission coefficients. (b) shows a zoomed view of the energy eigenvalues. (c) contains a comparison between the TB and the MFH models for a GQD consisting of 5,514 carbon atoms.}
  \end{center}
\end{figure}

To reveal the effect of the size quantization, we systematically study a series of the
pristine hexagonal GQDs consisting of up to 10,806 atoms (R = 10.4 nm).
After this limit, physical properties approach to those of the corresponding bulk material \cite{li2019review}.
A Coulomb potential is placed at the center of each hexagonal GQDs; see the inset of Fig. \ref{fig:fig_1}(a).
To discuss the size effect within the MFH model, energy eigenvalues of the lowest bound
states of all samples as a function of the coupling
strength $\beta$ and zoomed portion around the critical coupling constant
$\beta_{c}$ are shown in Fig. \ref{fig:fig_1}(a) and Fig. \ref{fig:fig_1}(b), respectively. In Fig. \ref{fig:fig_1}(a) and (b),
each of the lowest angular momentum channels is doubly degenerate due to the valley symmetry \cite{zarenia,sheng}.
In short, the spin and valley degeneracies are preserved as a function of the $\beta$. As a result, the MFH results do not make any
discrimination between the spin components due to the spin symmetry. From now on,
TB results are given by the black lines, while results of the
spin-up and the spin-down can be followed by the red and
blue lines in each of the remaining graphs, respectively.
Different kinds of symbols in Fig. \ref{fig:fig_1}(a) show the size of
the hexagonal GQDs, and we also use these symbols in the remaining part of the paper.

Each of the lowest bound states dives into the negative
energies at the same value of the coupling strength
that is 0.6. It can be accepted as a critical point at this stage,
and we will discuss this point in more detail below.
It is clear that effect of the size is negligible due to the special characteristic of
their band gaps. The collapse states are pinned at the
Dirac point (DP) as clearly shown in the experiments
\cite{wang2013observing,mao2016realization}.
In this sense, our results indicate that the zero
energy plays the same role with the DP in bulk
graphene. In contrast, Fermi level follows the highest
filled level due to a constant number of electron-like Dirac
fermions. Our results pave the way for the examination
of reconstruction of the Dirac vacuum within quite small
sample sizes by a low computational cost.

On the other hand, the electron-electron interactions
in half-filled MFH model are set by the on-site U and
off-site V terms as given in Eq. \ref{eq:eq1}.
Energy eigenvalues of TB model are compared
with those of the MFH model by setting the off-site term V to zero.
As it is clear from Fig. \ref{fig:fig_1}(c), the on-site term
U gives no contribution to the renormalization of the
$\beta_{c}$. In contrast, the off-site term V decreases
overscreening tendency
\cite{kotov2008electron,kotov2012electron}
of the TB approximation by smearing out the induced charge density
\cite{biswas2007coulomb}, and which turns out to
be a 20$\%$ increase in the $\beta_{c}$. It would be interesting to study screening properties in the GQDs as extensively
examined in bulk graphene \cite{biswas2007coulomb,shytov2007vacuum,gamayun2011magnetic,wang2012mapping,nishida2014vacuum,luican2014screening,wong2017spatially}, but we directly give a critical
bare valance charge Z$_{\text{c}}$
\begin{equation}\label{eq:eq7}
\text{Z$_{c}$}\left(\frac{2.2}{\kappa}\right) = \beta_{c}\qquad\rightarrow
\qquad
\text{Z$_{c}$} \approx 1.64 \text{,}
\end{equation}
where the dielectric constant $\kappa$ = 6, and the critical coupling constant $\beta_{c}$ equals to 0.6.
It indicates that impurities with the critical
valence charge Z$_{\text{c}}$ $\approx$ 1.64 can be used to create an
artificial supercritical nuclei for all GQD sizes. Our result is
also consistent with the previous one in which Z$_{\text{c}}$ is calculated to be larger than
unit charge \cite{terekhov2008screening}. The tight-binding
result for one particular hexagonal GQD consisting of
5,514 atoms shows (Fig. 1(c)) that the lowest bound state
enters the supercritical regime at $\beta_{\text{c}}$ = 0.5,
same as what is expected for bulk graphene. In compliance with our
results, the critical wave functions of the circular GQDs merge into negative energies
at the value of $\beta_{\text{c}}$ = 0.5 within the effective mass
approximation with an infinite mass boundary condition \cite{van2017graphene}.

The band gap in the GQDs is only due to size restriction of
massless Dirac fermions, and we give an interaction-induced
renormalization of the $\beta_{\text{c}}$. This gap should not be confused
with that of a gapped graphene monolayer \cite{zhou2007substrate},
modelled by adding a mass term in bulk graphene \cite{kotov2008polarization,chakraborty2013effect,kuleshov2015coulomb}. Reported values of the $\beta_{\text{c}}$ up to $\simeq$ 0.9
\cite{pereira2008supercritical,zhu2009electronic}
are calculated for the non-interacting massive
Dirac fermions, where the critical point is defined as the crossing of the
collapse state with the lower continuum \cite{pereira2008supercritical}, instead of DP in our calculations.
In addition, Fermi level moves automatically down due to the absence of charge compensation, similar to our case.
\subsection{\label{sec:level5} Transmission coefficients}
\begin{figure}[b]
\begin{center}
\includegraphics[width=\linewidth, height=4in,keepaspectratio]{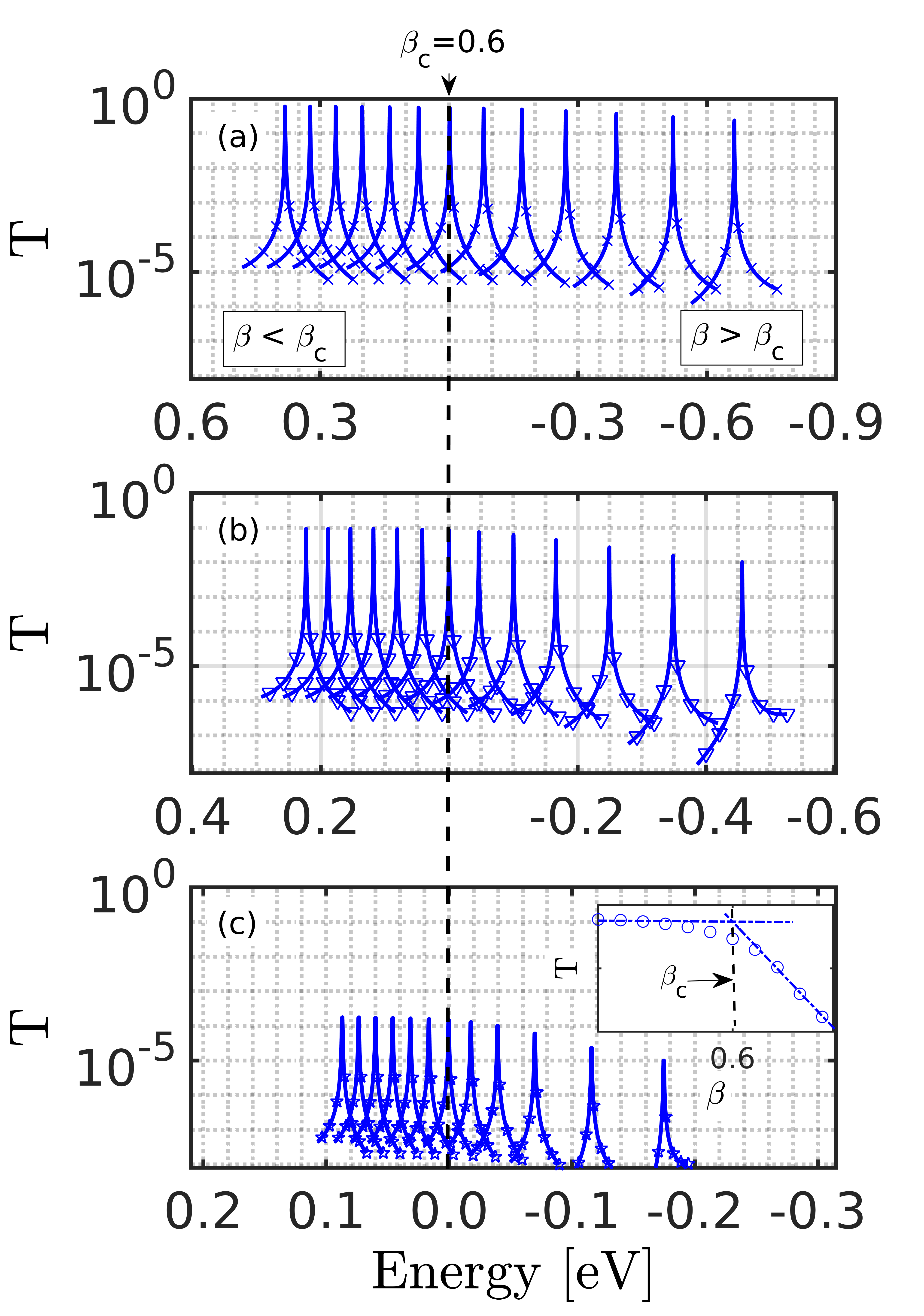}
\caption{\label{fig:fig_2} The transmission coefficients in (a), (b), and (c) for the number of 546,
1,626, and 10,806 atoms, respectively. The behaviour of
transmission coefficients obviously corresponds to two different regime.
Inset in (c): the critical coupling constant $\beta_{\text{c}}$ is at the point of
intersection of two lines on a linear scale.}
  \end{center}
\end{figure}

The transmission coefficients T of the lowest bound states as a function of the
energy E are shown in Fig. \ref{fig:fig_2}(a),
(b), and (c) for the hexagonal GQDs
consisting of 546, 1,626, and 10,806 atoms, respectively.

In all figures, from
left to right, each of the transmission peaks is
calculated for the consecutive values of the $\beta$ with a step size of 0.1,
starting from the $\beta$ = 0. When the subfigures are compared with
each other in the absence of the Coulomb potential, i.e., $\beta$ = 0, it
is clear that the transmission coefficients of the lowest
bound spin-down states decrease inversely with the size
of the GQDs and reaches its minimum for the GQD
that contains 10,806 atoms. It can be noted that the
maximum transmission is observed for the all GQDs
consisting of up to 222 atoms. For the subcritical range
0 $<$ $\beta$ $<$ $\beta_{\text{c}}$, the transmission coefficients do not make
significant changes. In other words,
the transmission coefficients of the lowest bound states remain almost
the same in the subcritical regime due to
the absence of the backscattering in the presence of the
central Coulomb impurity \cite{novikov2007elastic}.

When the coupling constant exceeds the critical value of
$\beta_{\text{c}}$ = 0.6, those coefficients drop immediately because
of the collapse of the wave functions. The peak values of the transmission coefficients are
plotted as a function of the coupling strength $\beta$ in the
inset of Fig. \ref{fig:fig_2}(c) for the GQD consisting of 10,806 atoms.
Two different regime are represented with the lines, and the
point of intersection clearly exhibits the $\beta_{\text{c}}$.
\section{\label{sec:level6} CHARGED VACANCY}
\subsection{\label{sec:level7} Spin and valley splittings}
\begin{figure}[htb]
\begin{center}
\includegraphics[width=\linewidth, height=4in,keepaspectratio]{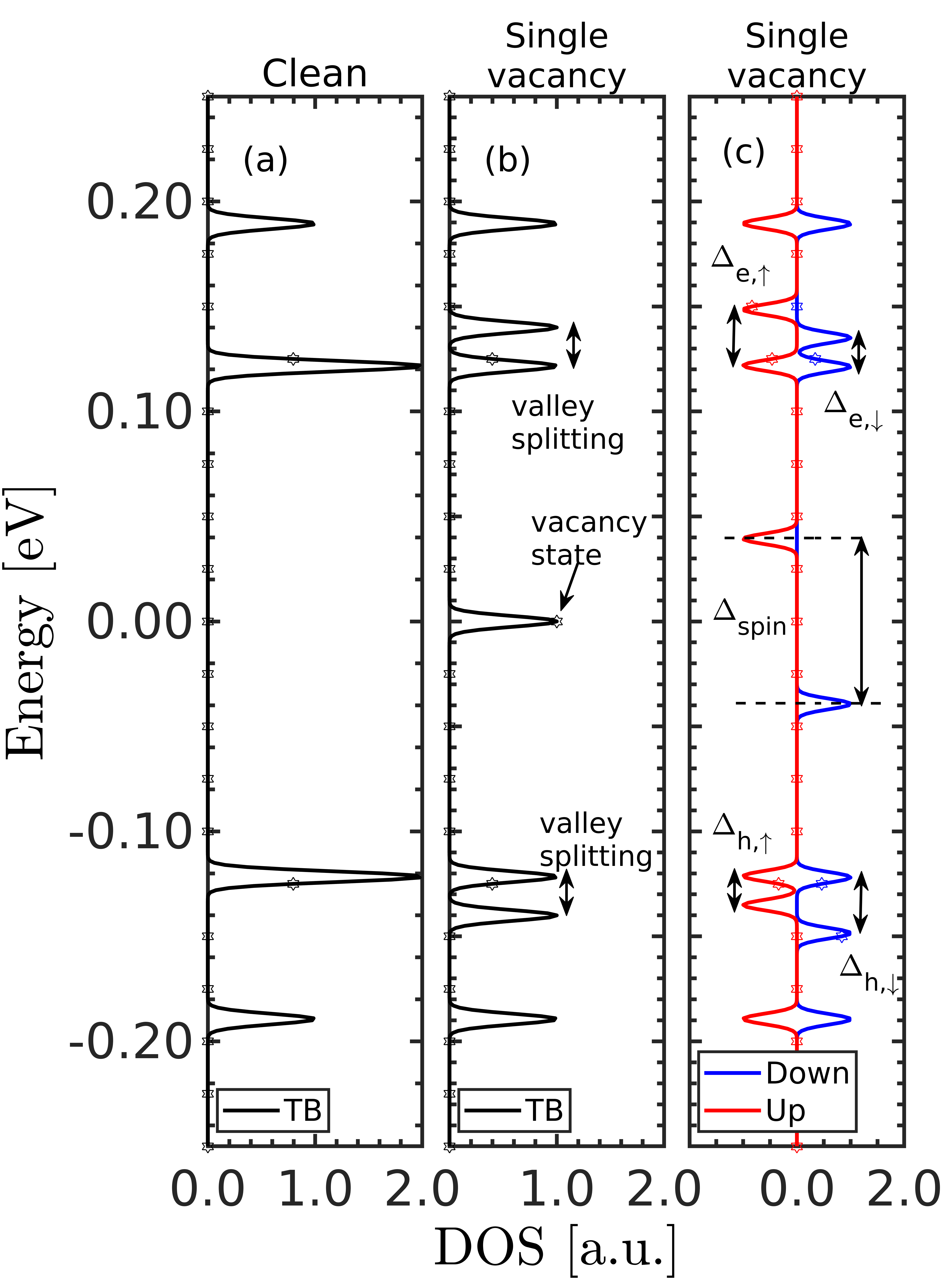}
\caption{\label{fig:fig_3} (a) total densities of states are shown for a pristine hexagonal GQD consisting of 5,514 atoms,
(b) total TB DOS belongs to the same GQD that contains a bare vacancy placed near the center, and
(c) contains the spin and valley splittings for both spin components.}
  \end{center}
\end{figure}

The breaking of the four-fold symmetry in nanographene and related
structures is a vital importance in understanding
the electronic as well as magnetic properties \cite{altintacs2018defect}. In this sense,
we analyze the sublattice-induced symmetry breaking staring from
the pristine hexagonal GQDs. DOS obtained for the clean hexagonal GQD consisting of 5,514
atoms using the TB model shows that the highest (lowest) occupied (unoccupied) state in the
valence (conduction) band is doubly degenerate (Fig. \ref{fig:fig_3}(a)). It can be noted that all sizes
have the same valley symmetry \cite{devrim}, and the valley degeneracy is observed
in both the TB and the MFH models in the same way.

When a single carbon defect is
intentionally created by removing of the $\pi_{z}$ orbital of the sublattice
A from the central benzene, the broken symmetry of the valley states shows itself
as the valley splittings with equal magnitude in electron and hole channels within the
TB method, see the vertical arrows in Fig. \ref{fig:fig_3}(b). At Fermi level, we have a pronounced vacancy peak
due to intervalley scattering caused by a bare carbon vacancy.

This vacancy state splits into up and down vacancy states
with equal spin probability and the occupation of $<$n$_{\downarrow}$$>$ = 1 and $<$n$_{\uparrow}$$>$ = 0
as shown in Fig. \ref{fig:fig_3}(c) when the interactions are turned on. These vacancy peaks are located symmetrically
with respect to Fermi level, and the spin splitting $\Delta_{\text{spin}}$ between them
is found to be 78 meV for this particular GQD. This splitting, also known as the spin polarization,
is proportional to the on-site Coulomb repulsion U \cite{gonzalez2016atomic}. When it comes to the valley splittings,
the picture becomes much more complicated. Note that the total DOS distribution of the spin-up contains two unequivalent
valley splittings. In the electron channel, we have the valley splitting $\Delta_{e,\uparrow}$ of 26 meV. In the hole channel,
the valley splitting $\Delta_{h,\uparrow}$ is found to be 13 meV. Similarly, the total DOS distribution of the spin-down
has two unequivalent valley splittings in both channels. Interestingly, there is an additional symmetry related to the valley
splittings dictated by the electron-hole symmetry. In the presence of a bare vacancy on the A sublattice, that is given by
\begin{equation}\label{eq:eq8}
\Delta_{\text{e},\downarrow} = \Delta_{\text{h},\uparrow} \text{,}
\qquad
\Delta_{\text{h},\downarrow} = \Delta_{\text{e},\uparrow} \text{.}
\end{equation}

\begin{figure}[t]
\begin{center}
\includegraphics[width=\linewidth, height=4in,keepaspectratio]{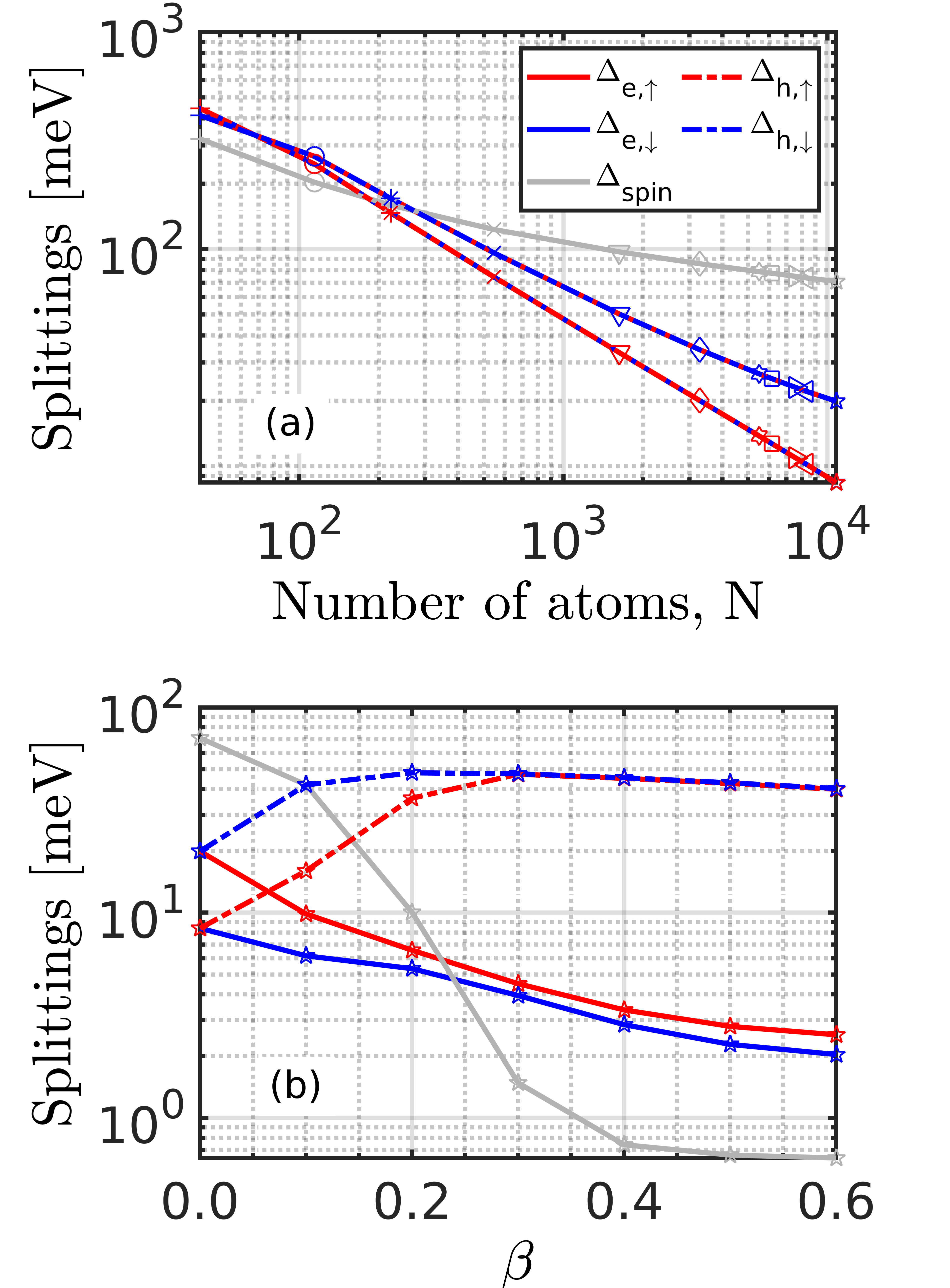}
\caption{\label{fig:fig_4} (a) clearly shows the spin and valley splitings as a function the size of the hexagonal GQDs.
Additional symmetry in Eq. 8 can be followed by the overlapped lines. (b) the spin splitting disappears as a function of $\beta$, while the valley splittings
do not completely vanish.}
  \end{center}
\end{figure}

As yet there is no discussion on the effect of the size on the splittings.
To analyze the size dependence, the spin and valley splittings
are plotted as a function of the size of
the hexagonal GQDs in Fig. \ref{fig:fig_4}(a). It is clear that the valley
splittings dominate the spin splitting at small sizes. On the contrary, for larger
sizes, the valley splittings are quite small as compared to the
spin splitting in the presence of a single bare vacancy. Moreover, the additional
symmetry between the valley splittings, given in Eq. \ref{eq:eq8},
is conserved as a function of the size.

When the vacancy is positively charged with the Coulomb potential, the spin splitting decreases as a function the
coupling strength $\beta$ as shown in Fig. \ref{fig:fig_4}(b). The quenching of the spin splitting occurs
at the coupling constant of $\beta$ = 0.4 that lies in the subcritical regime. It mimics that
the local magnetic moment can be tuned with the help of a charged vacancy.

The situation is totaly different in the valley splittings
depending on the occupation of the states. While the valley splittings of $\Delta_{h,\uparrow}$ and
$\Delta_{h,\downarrow}$ increase as a function of the coupling strength, both $\Delta_{e,\uparrow}$
and $\Delta_{e,\downarrow}$ show a decrement. However, all valley splittings continue to exist.
As it is clear, the spin splitting has a different behaviour from that of the valley splittings for a charged vacancy,
and which could prevent the valley states mixing with the spin states.
\subsection{\label{sec:level8} State characteristics}
\begin{figure}[b]
\begin{center}
\includegraphics[width=\linewidth, height=4in,keepaspectratio]{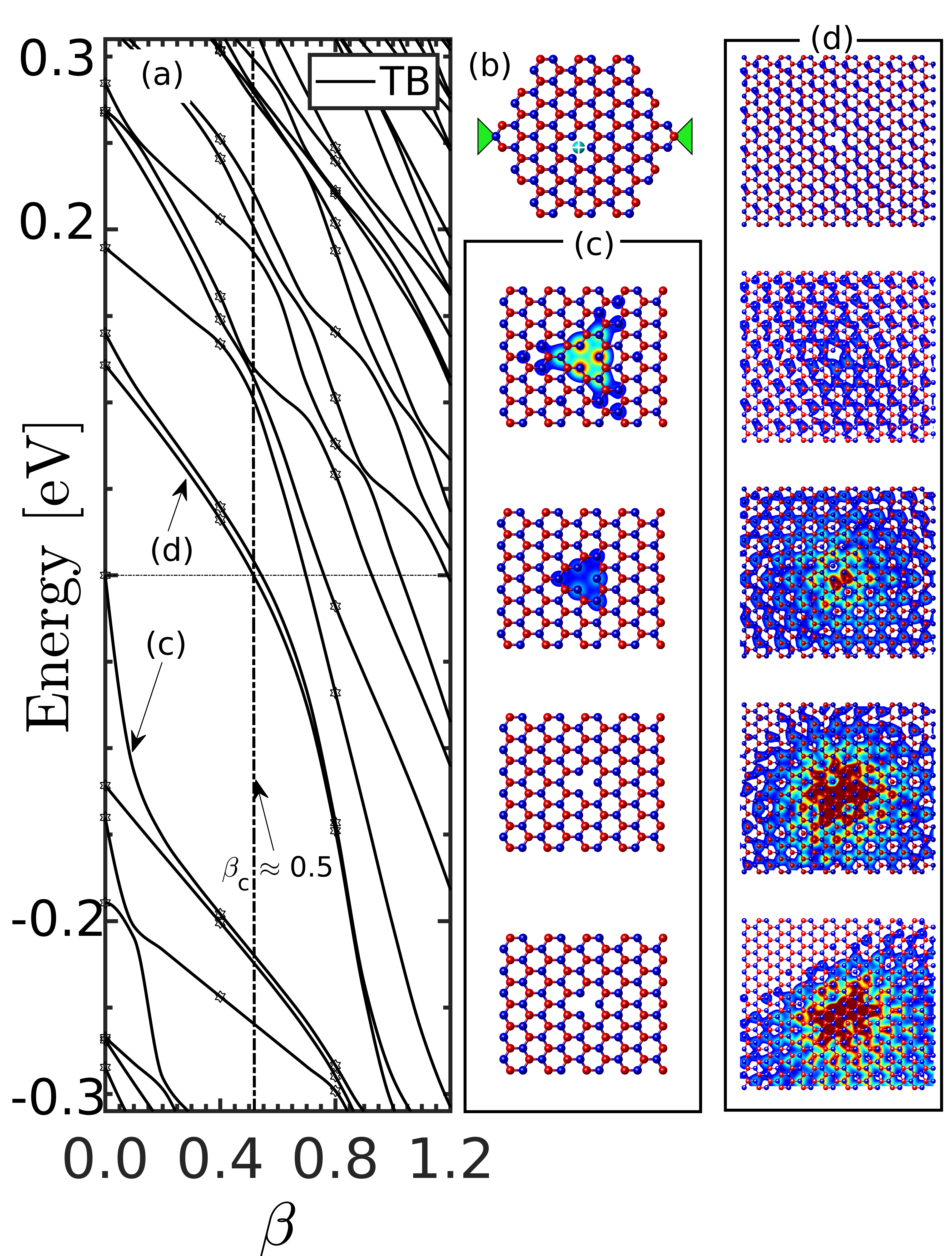}
\caption{\label{fig:fig_5} The energy spectrum of TB model as a function of the $\beta$ is shown in (a).
The positions of the leads and the bare carbon vacancy are sketched in (b).
Scaled electronic densities per lattice of the vacancy state, i.e., LDOS, for the $\beta$ = 0, 0.1, 0.2, and 0.3 can be seen in (c),
from top to bottom. The quasi-localization of the lowest bound
state is demonstrated in (d) for the $\beta$ = 0, 0.3, 0.6, 0.8, and 1.0, from top to bottom.}
  \end{center}
\end{figure}

The TB energy spectrum of a GQD consisting of 5513 atoms is plotted in Fig. \ref{fig:fig_5}(a)
as a function of the $\beta$. The vacancy state, labelled as (c) in Fig. \ref{fig:fig_5}(a), is pinned at
the energy origin and dives immediately into negative energies when the carbon vacancy is charged.
From top to bottom, the spatial distributions of the $\pi_{z}$-derived state are shown in Fig. \ref{fig:fig_5}(c)
for the following values of the $\beta$ = 0, 0.1, 0.2, and 0.3, respectively.
When we zoomed into the bare defect (at the top of Fig. \ref{fig:fig_5}(c)),
the triangular interference pattern due to intervalley
scattering can be seen as a characteristic spatial shape \cite{ugeda2010missing}.
As the $\beta$ is increased, the intervalley scattering is gradually surpassed by the intra$-$valley scattering,
and finally the uniform distribution of the vacancy state takes place at the $\beta$ = 0.2 and 0.3.
It means that highly localized defect state returns to its original bound state characteristic; however,
these scaled figures render the uniform spatial distribution invisible. This particular behaviour
will be strengthened below by means of the transmission coefficients.

It is also shown the spatial extension of the state labeling as (d) in Fig. \ref{fig:fig_5}(a). From top to bottom,
Fig. \ref{fig:fig_5}(d) exhibits the spatial extension of the critical state around the vacancy for
the $\beta$ = 0, 0.3, 0.6, 0.8, and 1.0, respectively. Uniform spatial
extension of the critical state exists for the $\beta = 0$
as shown at the top of Fig. \ref{fig:fig_5}(d). On exceeding the critical value, the $\beta$ $\gtrsim$ 0.5,
the critical state dives into negative energy spectrum, so that the appearance of the quasi-localized
state occurs around the charged vacancy. It is actually defined as the counterpart
of the 1S atomic collapse state in Ref. \cite{mao2016realization}.
\begin{figure}[b]
\begin{center}
\includegraphics[width=\linewidth, height=4in,keepaspectratio]{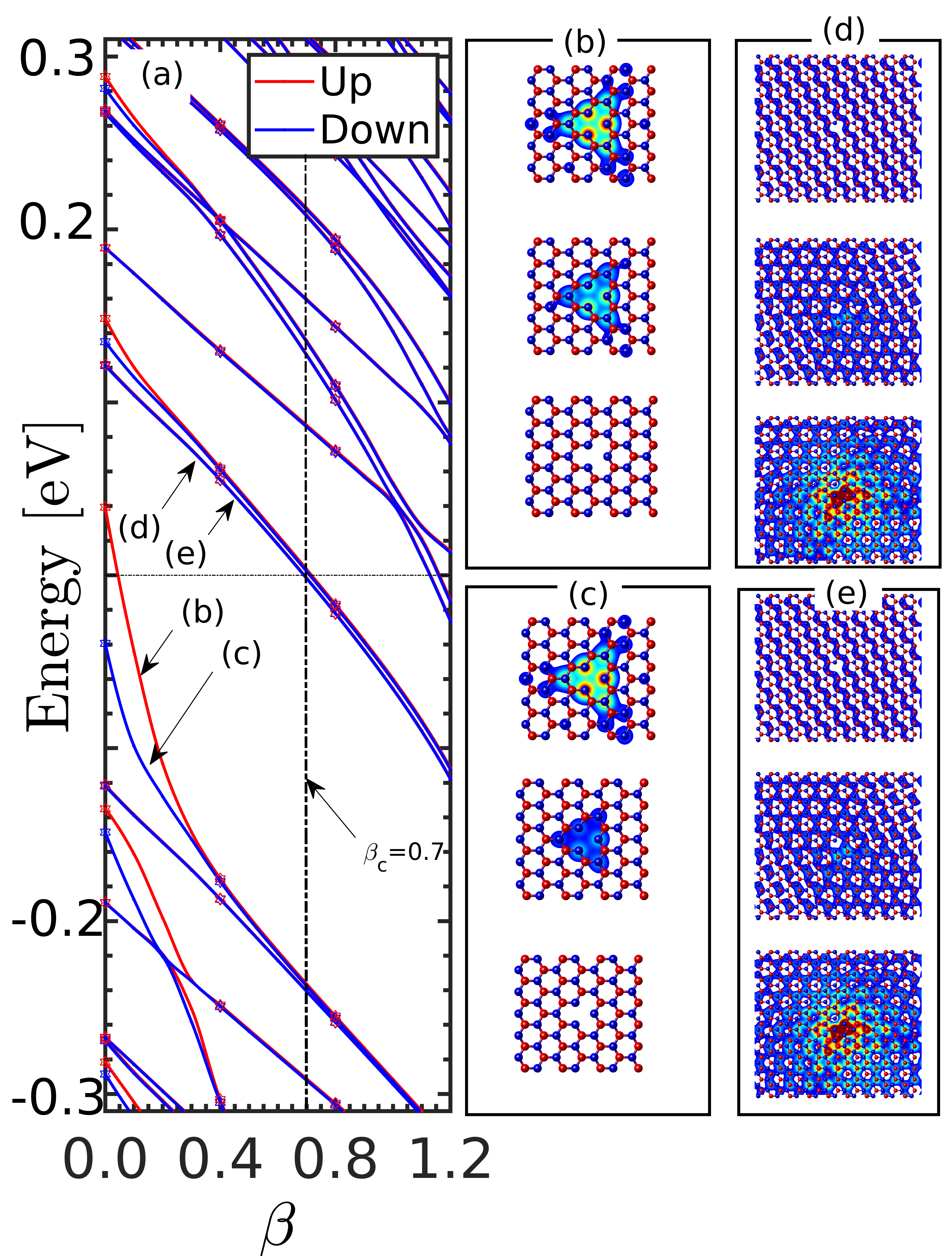}
\caption{\label{fig:fig_6} The energy spectrums of the spin-up and spin-down are shown in (a).
The $\beta_{c}$ equals to 0.7 for a Coulomb charged vacancy. Scaled electronic densities for the vacancy states can be seen in (b) and (c) for the $\beta$ = 0, 0.2, and 0.4,
from top to bottom. In (d) and (e), the behaviour of the critical states for the $\beta$ values of 0, 0.7, and 1.2
can be seen starting from top.}
  \end{center}
\end{figure}

When the electron interactions are turned on, we have a different picture. The energy spectrum of the
spin-up is superimposed to that of the spin-down in Fig. \ref{fig:fig_6}(a)
as a function of the $\beta$. If Fig. \ref{fig:fig_6}(a) is analyzed, the spin
symmetry does not exist up to the $\beta$ = 0.4. In other words, an exact overlap of the energy spectrums
occurs at the $\beta$ = 0.4 meaning that the spin symmetry is regained for the larger
coupling strength values as previously discussed.

There is a defect state in the spin-up spectrum that is labelled as (b)
in Fig. \ref{fig:fig_6}(a). Its spatial distribution is displayed in
Fig. \ref{fig:fig_6}(b) for the $\beta$ = 0, 0.2, and 0.4
starting from the top. The defect state in the spin-up spectrum merges into negative energies
when the $\beta$ exceeds 0.1. The ideal triangular interference pattern
characteristic starts to decay, indicating a uniformly distribution on the
lattice sites. On the other hand, the defect state in the spin-down spectrum loses its
triangular shape from the moment the vacancy begins to charge, and
similarly it has a uniform distribution at the $\beta$ = 0.4 as shown
at the bottom of Fig. \ref{fig:fig_6}(c). At a value of the $\beta$ = 0.7, both spectrums have new diving
levels; see in Fig. \ref{fig:fig_6}(a). Both of the critical states
become quasi-localized in the supercritical regime as displayed in
the right columns for spin-up (d) and spin-down (e) states for the
$\beta$ = 0, 0.7, and 1.2, from top to bottom, respectively. As compared to
the non-interacting case, the critical coupling constant is renormalized to
the $\beta_{c}$ = 0.7 in the presence of electron-electron interactions.
The critical states in both energy spectrums collapse at the same $\beta_{c}$.
The values of the $\beta_{c}$ are valid for
all sizes of the hexagonal GQDs when a vacancy is charged with the Coulomb potential.
\subsection{\label{sec:level9} Transmission coefficients and staggered magnetization}
\begin{figure}[b]
\begin{center}
\includegraphics[width=\linewidth, height=\linewidth,keepaspectratio]{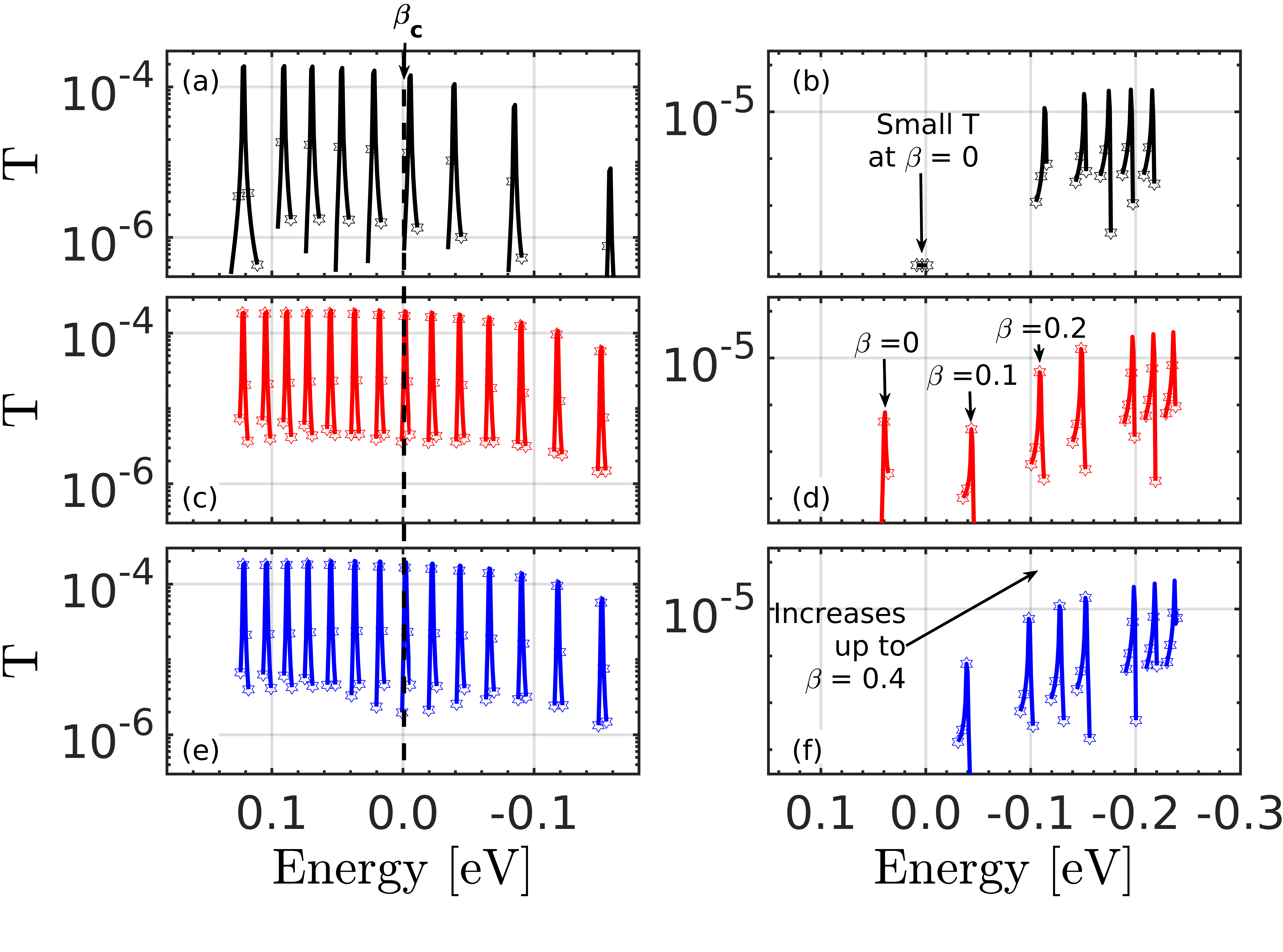}
\caption{\label{fig:fig_7} Transmission coefficients of the critical states of TB in (a), spin-up in (c), and
spin-down in (e) spectrums, while the vacancy states are given in (b), (d), and (f).}
  \end{center}
\end{figure}

Transmission coefficients of the critical states of TB,
spin-up, and spin-down spectrums are calculated.
First of all, in Fig. \ref{fig:fig_7}(a), (c), and (e),
the transmission coefficients are approximately 2 $\times$ 10$^{-4}$
in the subcritical regime $\beta$ $<$ $\beta_{c}$. It can be inferred that there is no
a direct effect of including electron-electron interactions on the transmission
coefficients. Whenever a critical state dives into the negative energies,
which happens at the $\beta_{c}$ $\approx$ 0.5 for TB and $\beta_{c}$ = 0.7 for
the MFH spectrums, the transmission coefficients immediately drop.
Basically, the quasi-localized character of these states is responsible
for a decrement observed in transmission coefficients.

The transmission coefficient of the vacancy state
in TB spectrum is plotted in Fig. \ref{fig:fig_7}(b).
It has too small transmission value at the $\beta$ = 0,
whereas the transmission coefficient increases and stays
nearly the same for the $\beta$ $>$ 0.1.

This result actually points out that returning to the bound state characteristic leads to an increase in
the transmission coefficient (see again Fig. \ref{fig:fig_5}(c)). The same physics is valid for
all the vacancy states observed within the MFH models. As shown in Fig. \ref{fig:fig_7}(d), the transmission
coefficient for the vacancy state in the spin-up spectrum reaches its maximum at the $\beta$ = 0.4,
although there is a small deviation at the $\beta$ = 0.1. When it comes to the vacancy state in the spin-down
spectrum, the transmission coefficient (Fig. \ref{fig:fig_7}(f)) gradually
increases up to the $\beta$ = 0.4 when we charge the defect.
The reason for this is the recovering of the initial bound state characteristic.

\begin{figure}[htb]
\begin{center}
\includegraphics[width=\linewidth, height=1.8in,keepaspectratio]{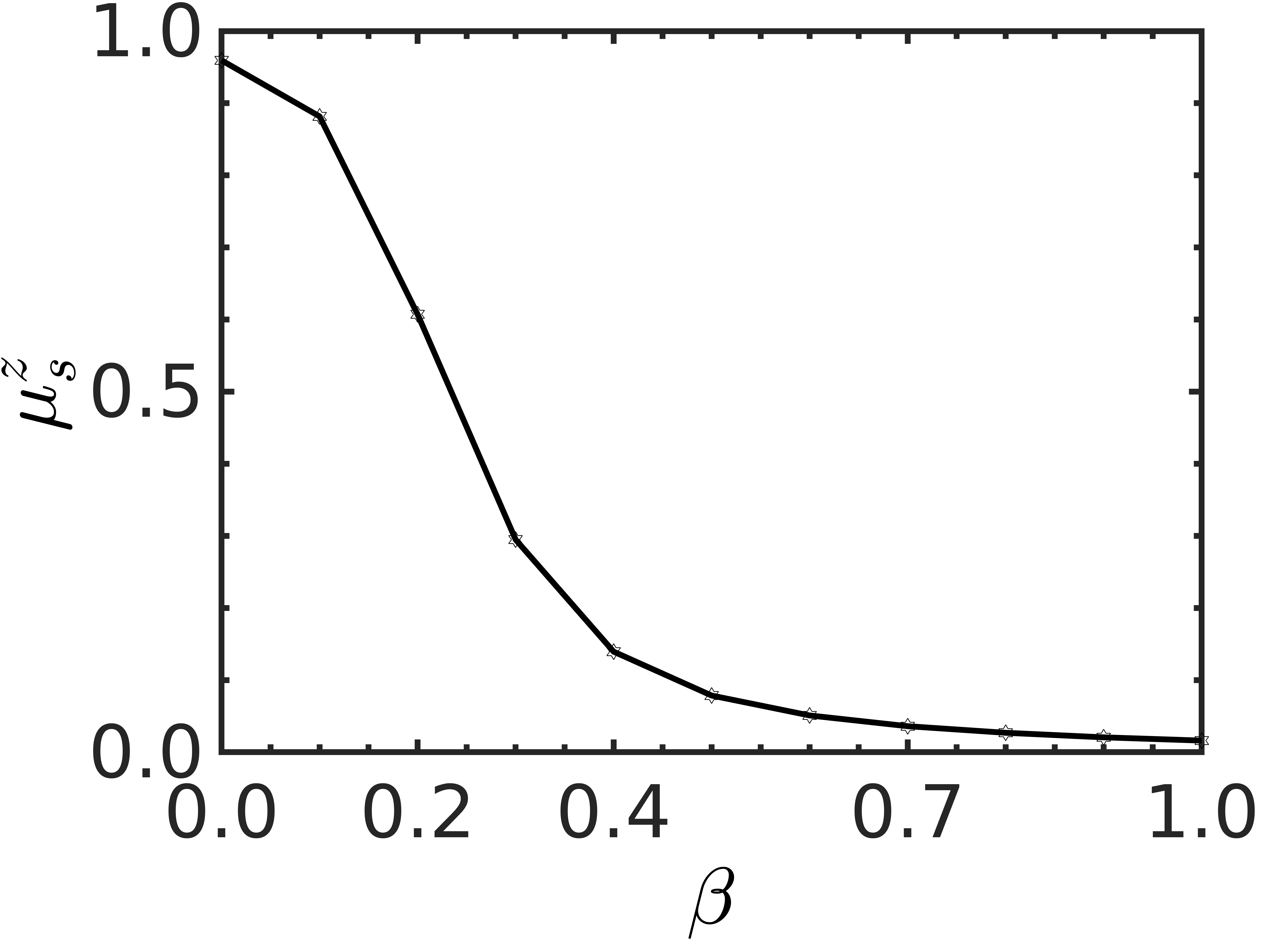}
\caption{\label{fig:fig_8} The quenching
of staggered magnetization $\mu_{s}^{z}$ is given as a function of the $\beta$. Up to the $\beta$ = 0.4,
a large portion of the $\mu_{s}^{z}$ disappears.}
  \end{center}
\end{figure}

As plotted in Fig. \ref{fig:fig_8}, a large amount of the staggered magnetization $\mu_{s}^{z}$ vanishes
when the coupling constant $\beta$ equals to 0.4. This behaviour guarantees that
the spin symmetry is regained for a Coulomb charged vacany. In this manner, the mechanisms of evolution, observed for
the vacancy states in Fig. \ref{fig:fig_6}(b) and (c), seem to be the underlying reason.
\section{\label{sec:level10} CONCLUSIONS}
In summary, the hexagonal GQDs with special characteristic band gap feature are studied by means of
TB and MFH models. The critical coupling constant is found to be $\beta_{c}$ = 0.5 for the non-interacting
cases when the Coulomb potential is placed at the center. However, the $\beta_{c}$ is renormalized to
0.6 for all sizes in the presence of off-site electron-electron interactions. It can be noted that the off-site
repulsion term is responsible for this incasement due to the long-range repulsive tail.
It is calculated that the central impurities with bare nuclear charge Z$_{\text{c}}$ $\approx$ 1.64 are at the edge of the supercritical threshold. Additionally, it is revealed that the transmission coefficients remain the same in the subcritical regime
$\beta$ $<$ $\beta_{c}$ due to the absence of the backscattering. However, those values in the supercritical regime
$\beta$ $>$ $\beta_{c}$ shows a strong dependence on the coupling strength.

It is revealed with the help of DOS that a bare vacancy gives rise to the simultaneous formation of the
valley and spin splittings. The spin splitting is larger than the valley splittings for the larger sizes,
whereas the valley splittings become dominant for the small sizes. As the coupling strentgh $\beta$ is
increased, the spin splitting vanishes at $\beta$ = 0.4. The behaviour of valley splittings completely depends
on the occupation of the valley states. In the hole (electron) channel, the valley splittings
show an increment (decrement) for the larger coupling strength. However, the valley splittings never vanish.
It signals that the mixing of the valley states with the spin states is not possible in the presence of a charged vacancy.

The formation of the quasi-localization
around a charged vacancy is monitored with the help of LDOS.
The critical state collapses when the coupling
constant exceeds $\beta_{c}$ $\approx$ 0.5
for TB and $\beta_{c}$ = 0.7 for the MFH models for a charged vacancy.
Furthermore, the transmission coefficient of the critical states decreases in the supercritical regime.
On the contrary, those coefficients of the vacancy states increase in the subcritical regime, as the coupling strength is increased.
The quenching of the spin splitting is also discussed with the help of the staggered magnetization
which reinforces the findings related to regaining of the spin symmetry.
\begin{acknowledgments}
We are grateful to A. M. Alt{\i}nta\c{s} and P. Hawrylak for stimulating discussions and to A. M. Alt{\i}nta\c{s} for helping with computational algorithms. This work was supported by
The Scientific and Technological Research Council of Turkey
(TUBITAK) under the 1001 Grant Project Number 116F152.
\end{acknowledgments}


\providecommand{\noopsort}[1]{}\providecommand{\singleletter}[1]{#1}%
\end{document}